\documentclass[a4paper,11pt]{article}

\usepackage[latin1]{inputenc}
\usepackage[top=2.5cm, left=2.5cm, right=2cm, bottom=2.5cm]{geometry}
\usepackage{graphicx}
\usepackage[centertags]{amsmath}
\usepackage{amsfonts}
\usepackage{amssymb}
\usepackage{amsthm}
\usepackage{newlfont}
\usepackage{xcolor}
\usepackage{soul}
\usepackage{lscape}
\usepackage{authblk}
\usepackage[mathscr]{eucal}

\usepackage{xcolor} \definecolor{darkgreen}{rgb}{0,.5,0}
\usepackage[colorlinks,filecolor=blue,citecolor=darkgreen,unicode]{hyperref}

\begin{document}
\title{Galilean Covariant Carroll--Field--Jackiw Electrodynamics}

\author[1]{H. Belich \thanks{humberto.belich@ufes.br}}
\author[2]{E. S. Santos \thanks{esdras.santos@ufba.br}}
\author[3]{C. Valcarcel \thanks{valcarcel.flores@gmail.com}}
\affil[1]{Centro de Ci\^encias Exatas, Universidade Federal do Esp\'irito Santo, $29.060-900$, Vit\'oria, ES, Brazil.}
\affil[2,3]{Instituto de F\'isica - Universidade Federal da Bahia, C\^ampus Universit\'ario de Ondina, $40210-340$, Salvador, B.A. Brazil}
\date{}

\maketitle

\begin{abstract}

We propose a non--relativistic version of the Carroll--Field--Jackiw theory in order to study the breaking of Galilean symmetry induced by the inclusion of an external tensor via Chern--Simons--like term in the Galilean covariant Lagrangian for the massive vector field. The results show that this model allows wave plane solutions with two frequency modes, i.e., it is possible to describe the phenomena of birefringence in the non--relativistic context. We also study the planar regime of this model in the two limits (electric and magnetic) of the usual electromagnetic field, obtaining the generation of topological mass and current of the Galilean fields. Finally, and following the same way, we propose a Podolsky electrodynamics with a Galilean--symmetry breaking term producing also the birefringence. 

\vspace{.5cm}


\end{abstract}

\section{Introduction}

The detection of the Higgs boson in 2013 by the Large Hadrons (LHC) collider consolidated the Weinberg-Salam-Glashow model (SM) \cite{Ali:1994wm}. Despite this great success, this model that aims to unify all known interactions, the gravitational interaction, is left aside. Therefore, it is necessary to search for a more fundamental theory, which includes gravitation.

How can we proceed with extending this model, which is fundamentally based on gauge and Lorentz symmetries to describe interactions? Electroweak unification is carried out by a complex scalar field that undergoes a spontaneous break of symmetry, generating mass to the Gauge bosons of the weak theory \cite{Ali:1994wm}. This mechanism, which had already been carried out similarly in superconductivity \cite{Anderson:1963pc}, became known as the Anderson--Higgs mechanism \cite{Higgs:1964pj}. 

The extension of this mechanism by replacing the role of the scalar fields
with another of higher rank spontaneously breaks the Lorentz Symmetry
\cite{belich_2007}. The discussion about the breaking of Lorentz symmetry started with Dirac \cite{Dirac1951}. Then Bondi and Gold showed that Dirac's behavior Electrodynamics, with spatial anisotropy, has cosmological implications \cite{Bondi1952}. In the 1980s, after the String Revolution, the idea of extending the Higgs Mechanism was presented and the Spontaneous Violation of Lorentz Symmetry (SVLS) was first proposed by Kosteleck\'y and Samuel \cite{Kostelecky:1988zi}. Taking into account the renormalizability \cite{Carroll:1989vb,Kostelecky:1999zh} these proposals were collected as the Standard Model Extension (SME) \cite{Colladay:1996iz,Colladay:1998fq}. By relaxing this condition, hence, we investigate possibilities out of SME (non--minimal SME), such as models with the Lorentz symmetry violation \cite{Belich:2004ng,Belich:2006tk,Belich:2006pi,Belich:2006wk,Belich:2007rg,Belich:2009gaj,Kostelecky:2009zp,Kostelecky:2011gq,Kostelecky:2013rta,Ding:2016lwt,Kostelecky:2018yfa,Kostelecky:2020hbb}. With the focus on geometric quantum phases, several works have dealt with
the Lorentz symmetry breaking effects through geometric quantum phases
\cite{Belich:2004ng,Belich:2009gaj,Silva:2013cpv,Ribeiro:2007fe,Belich:2011dz,Bakke:2013lqa,Bakke:2012gt,Belich_book}.

In parallel, topological materials where electronic transport takes place in $2+1$ dimensions, since the discovery of the Quantum Hall effect \cite{Haldane:1988zza}, have been the focus of much research today. Such solids which present the transport with an effective null fermion mass are known as Dirac materials, and graphene is an example where it occurs \cite{RevModPhys.81.109}. An interesting property that has been predicted in pyrochlore iridates \cite{PhysRevB.83.205101,PhysRevLett.108.046602} is known as the Weyl semi--metallic phase \cite{Shuichi_Murakami_2007}. In this new electronic state, the low--energy excitations are Weyl fermions \cite{rao2016weylsemimetalsshort,Jia_2016,RevModPhys.90.015001,QLu.2024}, and an effective field theory with Lorentz Violation Symmetry made by background fields has appeared naturally. 

Despite its relativistic character, the theories and models described above also present low--energy limits, which need a non--relativistic approach. In this sense, the pioneering works of Sen Gupta \cite{SenGupta:1966qer} and L\'evy--Leblond \cite{AIHPA_1965__3_1_1_0}, there has been explored several routes to build non--relativistic theories \footnote{Sen Gupta transformations can be used to map Lorentzian theories to the $c=0$ limit: Carrollian relativity \cite{Banerjee:2024jub}}. One of them is working non--covariantly \cite{Banerjee:2022eaj,Banerjee:2023wcp,Banerjee:2023koh}. Another one is to work in covariant form, using a five--dimensional spacetime \cite{Duval:1984cj}. This formalism allows us to build Galilean covariant field theories (see \cite{Santos:2004pq,Bagchi:2022twx}) and has been applied to different areas, for example: Scalars, Complex, and Galilean Electromagnetic fields \cite{Santos:2004pq}, Fermionic fields \cite{deMontigny:2007fqc}, spin--two fields \cite{Cuzinatto:2009aq}, Einstein's gravity \cite{Ulhoa:2009at}, teleparallel gravity \cite{Ulhoa:2011ei}, Bose--Einstein condensation \cite{ABREU2015612}, heavy mesons \cite{Abreu:2018ipp}, Galilean Duffin--Kemmer--Petiau fields in arbitrary dimensions \cite{deMontigny:2020evx} among others.  (see \cite{Bergshoeff:2022eog,Baiguera:2023fus}, for example),

In this paper, we build a Galilean covariant version of the  Carroll--Field--Jackiw (CFJ) theory \cite{Carroll:1989vb} in order to examine Galilean--symmetry breaking. To perform this approach, we introduce a Chern--Simons term in the Lagrangian of the vector field with a background tensor field, which breaks the Galilean symmetry and provides a dispersion relation with two different modes in the context of Galilean CFJ electrodynamics. In addition, we analyze the wave plane solutions for the dimensional reduced Galilean CFJ theory, in order to find topological masses that can couple with the mass of the Galilean fields. Finally, we add to the CFJ theory a higher derivative term and analyze the birefringence effect.

The paper is organized as follows. In Section \ref{gal_space} we review the Galilean spacetime and discuss how we build covariant theories. In section \ref{Gal_CFJ} we build the Galilean covariant Carroll--Field--Jackiw theory. This model depends on a constant background tensor. In Section \ref{Gal_disp} we obtain a plane--wave solution with two frequency modes for a specific choice of this background. In Section \ref{dim_red} we perform the dimensional reduction imposing a constraint on one spatial coordinate and analyze the classical solution. In Section \ref{Podolsky} we introduce higher derivative terms to the Carroll--Field--Jackiw theory and found a new dispersion relation with birefringence. Finally, in section \ref{final_remarks} we discuss our results.

\section{The Five--dimensional Galilean spacetime}\label{gal_space}

In order to define covariantly field theories invariant under the Galilean group, it is convenient to enlarge the spacetime and define a five--dimensional space, denoted by $\mathcal{G}_{(4,1)}$, with coordinates $x^\mu=\left(\mathbf{x},x^4,x^5\right)$. This space has a line element
\begin{equation}
\mathrm{d}s^{2}=\eta_{\mu\nu}\mathrm{d}x^\mu\mathrm{d}x^\nu  
\end{equation}
where the metric components are:
\begin{equation}\label{metric}
\eta_{\mu\nu}=\left(\begin{array}{ccc}
\mathbb{I}_{3\times3} & 0 & 0\\
0 & 0 & -1\\
0 & -1 & 0
\end{array}\right).    
\end{equation}
We define the inner product between two Galilean vectors as 
\begin{equation}
 x_\mu y^\mu=\eta_{\mu\nu}x^\mu y^\nu=\mathbf{x}\cdot\mathbf{y}-x^4 y^5-x^5 y^4.   
\end{equation}
In particular, the norm of a vector $x^\mu$ is $x^{2}=\mathbf{x}^{2}-2x^{4}x^{5}$. Due to the minus sign, the norm of a vector can be positive, negative, or zero. 

It can be shown that the line element is invariant under the following transformation
\begin{eqnarray}
x'^{i}	&=&	R_{\;j}^{i}x^{j}-\beta^{i}x^{4}\\
x'^{4}	&=&	x^{4}\\
x'^{5}	&=&	x^{5}-\beta_{j}R_{\;i}^{j}x^{i}+\frac{1}{2}\mathbf{\beta}^{2}x^{4}.    
\end{eqnarray}
where $R^i_{\;j}$ and $\beta^i$ are constant parameters.

There are different embeddings of Euclidean space into $\mathcal{G}_{(4,1)}$. For example, one embedding is given by
\begin{equation}
\mathfrak{I}:\mathbf{x}\rightarrow x=\left(\mathbf{x},x^{4}=\bar{c}t,x^{5}=\frac{\mathbf{x}^{2}}{2\bar{c}t}\right).    
\end{equation}
where $\bar{c}$ is a constant with units of velocity, such that $t$ has units of time. We can set this constant to unit $\bar{c}=1$ to avoid clutter. In this embedding, to every Euclidean vector $\mathbf{x}$ corresponds a null vector in $\mathcal{G}$: $x^{2}=\mathbf{x}^{2}-2t\left(\frac{\mathbf{x}^{2}}{2t}\right)=0$.

Note that, performing the identifications: $x^{4}=\bar{c}t$, $\beta^{i}=\frac{u^{i}}{\bar{c}}$, and $x^{5}=\frac{\mathbf{x}^{2}}{2\bar{c}t}$, we obtain the Galilean transformation:
\begin{equation}
\mathbf{x}'=R\mathbf{x}-\mathbf{u}t,\qquad t'=t  
\end{equation}
where $R$ represents a rotation and $\mathbf{u}$ a Galilean boost. In order to avoid clutter, from now on we set $\bar{c}=1$.

A free Galilean particle with energy $E$, momentum $\mathbf{p}$ and mass $m$, has to satisfy the on--shell condition $\mathbf{p}^2-2mE=0$. This allows us to write the five--dimensional momentum as $p_\mu=(\mathbf{p},E,m)$. Then, it has a modulo $p_\mu p^\mu = \mathbf{p}^2-2Em=0$. We can uplift this condition and consider $p_\mu p^\mu = \mathbf{p}^2-2Em=-\kappa^2$, where $\kappa$ is a constant. Note that in the rest frame, $\mathbf{p}=0$, the particle has rest energy $E=\kappa^2/2m$. A detailed study of the Galilean particle can be found in \cite{Figueroa-OFarrill:2024ocf}.

\subsection{Dynamics of fields on Galilean space--time }

In \cite{Santos:2004pq} was shown that we can build a Galilean covariant theory by writing the Lorentzian version in the five--dimensional $\mathcal{G}_{(4,1)}$ space. For example, let us consider the five--dimensional scalar field action
\begin{equation}\label{scalar01}
    I_{\mathrm{scalar}} = \int \mathrm{d}^5x\,\left( -\frac{1}{2} \partial_\mu \phi \partial^\mu \phi - \frac{1}{2} \kappa^2 \phi^2\right)
\end{equation}
where $\phi$ is the scalar field and the indices are raised and lowered with the Galilean metric \eqref{metric}. The first term is the kinetic term and the second one is a Proca term with a coupling constant $\kappa$. The equation of motion is
\begin{equation}\label{scalar02}
    (\partial_\mu \partial^\mu - \kappa^2) \phi = (\nabla^2 - 2 \partial_4 \partial_5 - \kappa^2)\phi = 0.
\end{equation}
Note that, written covariantly, the equation of motion looks exactly like the Lorentzian counterpart. However, when we write the partial derivatives explicitly with the Galilean metric, the equation takes a different form. 

We can now find solutions to the equations of motion. First, consider $x^4$ to be the temporal coordinate and suppose that the scalar field does not depend on $x^5$. In this case, equation \eqref{scalar02} reduces to the Helmholtz equation:
\begin{equation}\label{scalar03}
\left(\nabla^2 - \kappa^2\right)\phi(\mathbf{x},t) = 0.
\end{equation}
On the other hand, if the fields depend on $x^5$, we can write\footnote{This is, in fact, a short notation for $\phi(\mathbf{x},t)\exp{(imx^5)}+\phi^*(\mathbf{x},t)\exp{(-imx^5)}$, where $\phi^*$ is the complex conjugated, the complete expression is needed to obtain a real field. However, there is no lost of physical information if we consider \eqref{ansatz}}
\begin{equation}\label{ansatz}
\phi(x^\mu) =\phi(\mathbf{x},t)\exp{(imx^5)}.    
\end{equation}
This is possible because we previously identified $p_5=m$ as the conjugated momentum of the coordinate $x^5$. In this case the equation \eqref{scalar02} reduces to
\begin{equation}
 i \partial_t\phi = \left(\frac{\nabla^2}{2m} - \frac{\kappa^2}{2m}\right)\phi .
\end{equation}
This is a Schroedinger equation for the $\phi$ field. Note that, in contrast with the relativistic theories, $\kappa$ is not the mass of the scalar field.
Another ansatz is to choose $\phi(x^\mu)=\phi(\mathbf{x},t)\exp{(\alpha x^5)}$, where $\alpha$ is a constant. In this case, the equations of motion reduce to the Heat equation. 

In this work, we are interested in solutions of the form \eqref{ansatz} since they have dynamics and non--relativistic plane wave solutions.

\section{The Galilean Covariant Carroll--Field--Jackiw theory}\label{Gal_CFJ}

In \cite{Carroll:1989vb}, Carroll, Field and Jackiw proposed the following relativistic Lorentz--violating model
\begin{equation}\label{CFJ}
I_{\mathrm{CFJ}} = \int_{\mathcal M} \mathrm{d}^4x\;\left[-\frac{1}{4}F_{\mu\nu}F^{\mu\nu}-\frac{1}{4}\epsilon^{\alpha\beta\mu\nu}v_\alpha A_\beta F_{\mu\nu}\right]    
\end{equation}
where $\mathcal M$ is the usual Minkowski spacetime, $A_\mu$ is the gauge field, $F_{\mu\nu}$ its corresponding field strength,  $\epsilon^{\alpha\beta\mu\nu}$ is the Levi--Civita symbol and $v_\mu$ is a constant four--vector, which produces the Lorentz symmetry violation. The first term is the usual Maxwell lagrangian and the second term is called Lorentz--symmetry breaking term. For a review of Lorentz--violating theories, see \cite{Belich_book,Mariz:2022oib}.

Now, let us generalize this model to a Galilean covariant theory: First, our manifold is now $\mathcal G$, the five--dimensional Galilean space discussed in the previous section. We upgrade $A_\mu$ to a five dimensional gauge field: $\mathcal{A}_{\mu}$, whose components are given by
\begin{equation}
\mathcal{A}_{\mu}=\left(\mathbf{A},-\phi_{\mathrm{m}},-\phi_{\mathrm{e}}\right)    
\end{equation}
where $\mathbf{A}$ is the vector potential and $\left(\phi_{\mathrm{e}},\phi_{\mathrm{m}}\right)$ are the electric and magnetic potentials, respectively \cite{Santos:2004pq}. We must stress that in the five--dimensional theory, the electric and magnetic potentials do not coexist. In the so--called magnetic limit we take $\phi_{\mathrm{e}}=0$, while in the electric limit we set $\phi_{\mathrm{m}}=0$.

The field strength of the gauge field is defined as usual: $\mathcal{F}_{\mu\nu}\equiv\partial_{\mu}\mathcal{A}_{\nu}-\partial_{\nu}\mathcal{A}_{\mu}$. It now contains ten independent components. We can write  
\begin{equation}
\mathcal{F}_{ij}=\epsilon_{ijk}B_{k},\qquad\mathcal{F}_{i4}=E_{i}^{\mathrm{m}},\qquad\mathcal{F}_{i5}=E_{i}^{\mathrm{e}},\qquad\mathcal{F}_{45}=-\partial_{4}\phi_{\mathrm{e}}+\partial_{5}\phi_{\mathrm{m}}.    
\end{equation}
where $\mathbf{B}$ is the magnetic vector field and $(\mathbf{E}^{\mathrm{e}},\mathbf{E}^{\mathrm{m}})$ are the electric and magnetic limits of the electric vector field.

In order to write the Lorentz--violating term in the Galilean context, we must upgrade the four--dimensional Levi--Civita symbol to a five--dimensional one. Then, when we contract it with the gauge and field strength, we still have two free indices to perform a contraction: $\epsilon^{\bar{\alpha}\bar{\beta}\sigma\mu\mu}\mathcal A_{\sigma}\mathcal F_{\mu\nu}$. Then, instead of using a four--dimensional constant vector, as in the relativistic case, we need to use a five--dimensional constant antisymmetric tensor $v_{\mu\nu}=-v_{\nu\mu}$. 

We have now the all elements to propose the Galilean version of the Carroll--Field--Jackiw theory, which action is given by
\begin{equation}\label{action}
I_{\mathrm{Gal-CFJ}}=\int_{\mathcal{G}}\mathrm{d}^{5}x\;\left[-\frac{1}{4}\mathcal{F}_{\mu\nu}\mathcal{F}^{\mu\nu}-\frac{1}{2}\kappa^2\mathcal{A}_{\mu}\mathcal{A}^{\mu}-\frac{1}{8}\epsilon^{\alpha\beta\sigma\mu\nu}v_{\alpha\beta}\mathcal{A}_{\sigma}\mathcal{F}_{\mu\nu}\right].
\end{equation}
The first term is the usual Maxwell Lagrangian and the second term is a Proca term with coupling constant $\kappa$. The third term is called the Galilean symmetry--breaking term, in analogy with the relativistic case. We rise and lower the indices with the Galilean metric \eqref{metric} and, by convention, $\epsilon_{12345}=1$. 

The equation of motion is
\begin{equation}\label{eom}
\partial_{\mu}\mathcal{F}^{\mu\nu}-\kappa^{2}\mathcal{A}^{\nu}-\frac{1}{4}\epsilon^{\alpha\beta\rho\sigma\nu}v_{\alpha\beta}\mathcal{F}_{\rho\sigma}=0.
\end{equation}
Let us define 
\begin{equation}\label{current}
\mathcal{J}^{\nu}\equiv\frac{1}{4}\epsilon^{\alpha\beta\rho\sigma\nu}v_{\alpha\beta}\mathcal{F}_{\rho\sigma}.    
\end{equation}
It is easy to check that this quantity is divergenceless  $\partial_{\nu}\mathcal{J}^{\nu}=0$. Applying the operator $\partial_\nu$ on the equation of motion, we obtain the Lorentz condition: $\partial_{\mu}\mathcal{A}^{\mu}=0$. Under this condition the equation \eqref{eom} reduces to
\begin{equation}\label{eom2}
\left(\partial_{\mu}\partial^{\mu}-\kappa^{2}\right)\mathcal{A}^{\nu}=\mathcal{J}^{\nu}.    
\end{equation}
The form of this equation is very suggestive; on the left side we have a Klein--Gordon type equation for the gauge field, and on the right side the quantity $\mathcal J^\mu$ can be interpreted as a conserved current. 

In components, the equations of motion take the form:
\begin{eqnarray}
\left(\nabla^{2}-2\partial_{4}\partial_{5}-\kappa^{2}\right)\phi_{\mathrm{m}}	&=&	\tilde{\mathbf{w}}\cdot\mathbf{B}-\mathbf{v}\cdot\mathbf{E}^{\mathrm{m}}\\
\left(\nabla^{2}-2\partial_{4}\partial_{5}-\kappa^{2}\right)\phi_{\mathrm{e}}	&=&	-\mathbf{w}\cdot\mathbf{B}+\mathbf{v}\cdot\mathbf{E}^{\mathrm{e}}\\
\left(\nabla^{2}-2\partial_{4}\partial_{5}-\kappa^{2}\right)\mathbf{A}	&=&	n\mathbf{B}+\frac{1}{2}\mathbf{v}\left(\partial_{4}\phi_{\mathrm{e}}-\partial_{5}\phi_{\mathrm{m}}\right)+\mathbf{w}\times\mathbf{E}^{\mathrm{m}}-\tilde{\mathbf{w}}\times\mathbf{E}^{\mathrm{e}}
\end{eqnarray}
where we have written the ten independent parameters from the background field tensor as three independent three--dimensional vectors $\mathbf{w}$, $\tilde{\mathbf{w}}$, $\mathbf{v}$  plus a scalar $n$:
\begin{equation}\label{parameters}
w_{i}\equiv v_{5i},\quad\tilde{w}_{i}\equiv v_{4i},\quad v_{ij}\equiv\epsilon_{ijk}v_{k},\quad n\equiv v_{54}.   
\end{equation}
Solutions of these equations of motion are generally difficult to find. However, we can choose the background vectors conveniently in order to find physically interesting solutions.

\section{Dispersion relations for the Galilean covariant CFJ theory}\label{Gal_disp}

It is known that the Lorentzian CFJ theory \eqref{CFJ} has wave solutions with dispersion relations with two modes. Our objective is to find a non--relativistic dispersion relation in the Galilean covariant theory. 

First, set all background vector parameters to zero $\mathbf{w}=\tilde{\mathbf{w}}=\mathbf{v}=\mathbf{0}$. We leave only the scalar field $n$. Furthermore, from now on we will assume that all fields depend on the $x^5$ coordinate as in \eqref{ansatz}. The equation of motion for the magnetic and electric potentials are
\begin{equation}
\left(\nabla^{2}+2im\partial_t-\kappa^{2}\right)\phi=0.
\end{equation}
where $\phi=(\phi_\mathrm{e},\phi_\mathrm{m})$. Let us perform a Fourier decomposition of the form
\begin{equation}\label{Fourier}
\phi\left(\mathbf{x},t\right)=\int\frac{\mathrm{d}^{4}k}{\left(2\pi\right)^{4}} \exp{(-i\mathbf{k}\cdot\mathbf{x}-i\omega t)}\,\hat\phi\left(\mathbf{k},\omega\right) 
\end{equation}
being $\mathbf{k}$ the wave--vector and $\omega$ the angular frequency and $\hat\phi$ the Fourier transform of the electric or magnetic potentials. Replacing in the equation of motion, we obtain the following dispersion relation:
\begin{equation}\label{non_rel_dis}
    \omega = \frac{\mathbf{k}^{2}}{2m}+\frac{\kappa^{2}}{2m}.
\end{equation}
This is the usual non--relativistic dispersion relation for Galilean fields. Since the angular frequency is related to Energy and the wave--vector to momentum, we notice that the term proportional to $\kappa^2$ in \eqref{non_rel_dis} is related to rest energy. 

Let us now check in the direction of the vector field, which satisfies the following equation
\begin{equation}
\left(\nabla^{2}+2im\partial_t-\kappa^{2}\right)\mathbf{A} = n \nabla\times \mathbf{A}.
\end{equation}
Again, performing a Fourier decomposition, we obtain the following relation
\begin{equation}
\left(2m\omega-\mathbf{k}^2-\kappa^{2}\right)\hat{\mathbf{A}} + in \mathbf{k}\times \hat{\mathbf{A}} = 0
\end{equation}
where $\hat{\mathbf A}$ is the Fourier transform. Furthermore, since we are looking at the spatial sector, the Lorentz condition reads $\mathbf{k}\cdot\hat{\mathbf{A}}=0$. After some vector algebra, the above  we obtain
\begin{equation}
\left[\left(2m\omega-\mathbf{k}^{2}-\kappa^{2}\right)^{2}-n^{2}\mathbf{k}^{2}\right]\hat{\mathbf{A}}  = 0.
\end{equation}
Since $\hat{\mathbf{A}}$ is arbitrary, the term between brackets must be zero. This is a quadratic equation for the angular frequency, in contrast with the quartic equation found in the relativistic case. We obtain the following dispersion relation:
\begin{equation}\label{birefri}
\omega=\frac{1}{2m}\left(\mathbf{k}^{2}+\kappa^{2}\pm n\left|\mathbf{k}\right|\right). 
\end{equation}
This is a non--relativistic dispersion relation with two frequency modes which depends on $n$ the only non--zero component of the background. This is clearly an effect of birefringence in vacuum, in analogy with the relativistic CFJ model. Furthermore, since the frequency is always real, the system is stable, at least for the special choice of background \eqref{parameters}.

\section{Dimensional reduction:}\label{dim_red}

The five-dimensional theory discussed above has three spatial components. However, there are situations in which we define a theory in a plane, i.e, with only two spatial components. The dimensional reduction of the relativistic CFJ theory was performed in \cite{Ferreira:2019ygi}, \cite{Belich:2002vd} . In this section we follow a similar procedure for the Galilean covariant CFJ model. 

Let us first consider that the fields do not depend on the spatial coordinate $x^{3}$. This means that
\begin{equation}
x^{\mu}=\left(\mathbf{x},x^{4},x^{5}\right),\qquad\mathbf{x}=\left(x^{1},x^{2}\right).    
\end{equation}
Furthermore, there is no $\mu=3$ components for the gauge field. 
This implies that
\begin{equation}
\mathcal{A}_3=0,\qquad  \mathcal{F}_{3\nu}=0.
\end{equation}
For the background tensor we choose
\begin{equation}
v_{53}=2\mu\quad v_{43}=2\tilde{\mu}    
\end{equation}
where $\mu$ and $\tilde{\mu}$ are constants. Replacing these conditions in \eqref{action} we obtain the dimensional reduced action
\begin{equation}\label{dr01}
I_{\mathrm{red}}= \int_{\mathcal G}\mathrm{d}^4x\;\left[	-\frac{1}{4}\mathcal F_{\alpha\beta}\mathcal F^{\alpha\beta}-\frac{1}{2}\kappa^{2}\mathcal A_{\alpha}\mathcal A^{\alpha}+\frac{1}{2}\epsilon^{5\sigma\alpha\beta}\mu \mathcal A_{\sigma} \mathcal F_{\alpha\beta}+\frac{1}{2}\epsilon^{4\sigma\alpha\beta}\tilde{\mu} \mathcal A_{\sigma}\mathcal F_{\alpha\beta}\right].    
\end{equation}
The first two terms represent the Maxwell and Proca terms, while the last two terms are the Chern--Simons terms, where $\mu$ and $\tilde\mu$ play the role of couplings of the Chern--Simons actions.

The equation of motion of the dimensional reduced model is given by
\begin{equation}\label{dr02}
\left(\partial_{\mu}\partial^{\mu}-\kappa^{2}\right)\mathcal A^{\nu} = -2\mu \mathcal F^{\nu}-2\tilde{\mu}\tilde{\mathcal F}^{\nu}
\end{equation}
where the dual vectors $\left(\mathcal F^\nu,\tilde{\mathcal F}^\nu\right)$ are defined by
\begin{equation}
\mathcal F^\alpha \equiv \frac{1}{2}\epsilon^{5\alpha\mu\nu} \mathcal F_{\mu\nu},\qquad \tilde{\mathcal F}^\alpha \equiv \frac{1}{2}\epsilon^{4\alpha\mu\nu} \mathcal F_{\mu\nu}.
\end{equation}
Since $\partial_\alpha \mathcal F^\alpha=\partial_\alpha \tilde{\mathcal F}^\alpha=0$, we can define a divergenceless current: $\mathcal J^\nu = -2\mu \mathcal F^{\nu}-2\tilde{\mu}\tilde{\mathcal F}^{\nu}$. In this case, the current has a topological origin, since they come from Chern--Simons terms. Furthermore, we can show that the non--zero components of the dual vectors are:
\begin{equation}
	-\tilde{\mathcal F}^{5} = \mathcal F^{4}= \frac{1}{2}\epsilon^{ij}\mathcal F_{ij}=B,\qquad
	F^{i}	=\epsilon^{ij}\mathcal F_{j4}=\epsilon^{ij}E_{j}^{\mathrm{m}},\qquad \tilde{\mathcal F}^{i}	=-\epsilon^{ij}\mathcal F_{j5}=-\epsilon^{ij}E_{j}^{\mathrm{e}}. 
\end{equation}
where $B$ is the magnetic field and $\mathbf{E}^{\mathrm{m,e}}$ is the electric vector. Note that in planar dynamics, the magnetic field has a single component.

Now, let us assume that the gauge field has a dependence on the fifth--coordinate as 
\begin{equation}
\mathcal A_\mu(x) = \mathcal A_\mu(\mathbf{x},t) \;\exp{(imx^5)}.  
\end{equation}
Then, the equation of motion can be written as Schrodinger equation, that is
\begin{equation}
	\left(\frac{1}{2m}\nabla^2-i\partial_t-\frac{\kappa^{2}}{2m}\right)\mathcal A^{\nu}=-\frac{\tilde{\mu}}{m} \tilde{\mathcal F}^{\nu}-\frac{\mu}{m} \mathcal F^{\nu}    \label{em-4} 
\end{equation}
where the term $-\frac{\tilde{\mu}}{m} \tilde{\mathcal F}^{\alpha}-\frac{\mu}{m} \mathcal F^{\alpha}$ plays the role of a Schrodinger current. If we take the different values for $\nu$ in the equation above, we have
\begin{eqnarray}
	\left(\frac{1}{2m}\nabla^2-i\partial_t-\frac{\kappa^{2}}{2m}\right)\mathbf{A}&=&-\frac{\tilde{\mu}}{m} \mathbf{E}^{\mathrm{e}}_\bot-\frac{\mu}{m} \mathbf{E}^{\mathrm{m}}_\bot,    \label{em-5} \\
	\left(\frac{1}{2m}\nabla^2-i\partial_t-\frac{\kappa^{2}}{2m}\right)\phi_{\mathrm{e}}&=&-\frac{\mu}{m} B,    \label{em-6} \\
	\left(\frac{1}{2m}\nabla^2-i\partial_t-\frac{\kappa^{2}}{2m}\right)\phi_{\mathrm{m}}&=&\frac{\tilde{\mu}}{m}B.    \label{em-7} 
\end{eqnarray}
where $\mathbf{E}_\bot = (-E_2,E_1)$ is the transverse electric vector field. Note that the equations for the potentials have plane--wave solutions only if the magnetic field is zero. 

Let us now examine equations (\ref{em-5})-(\ref{em-7}) in two scenarios: first, when $\tilde{\mu}=0$ and $\mu\neq 0$; followed by $\tilde{\mu}\neq 0$ and $\mu= 0$.
 
\subsection{First Case: $\tilde{\mu} =0$ and $\mu \neq 0$}
 
 If we take the electric limit, $\phi_{\mathrm{m}}=0$, the last equation becomes an identity, while the equations (\ref{em-5}) and (\ref{em-6}) are rewritten by the form
\begin{eqnarray}
\left(\frac{1}{2m}\nabla^2-i\partial_t-\frac{\kappa^{2}}{2m}\right)\mathbf{A}&=&-\frac{\mu}{m} \mathbf{E}^{\mathrm{m}}_\bot,    \label{em-8} \\
   \left(\frac{1}{2m}\nabla^2-i\partial_t-\frac{\kappa^{2}}{2m}\right)\phi_{\mathrm{e}}&=&-\frac{\mu}{m} B.   \label{em-9} 
 \end{eqnarray}
 Observing these equations we can see that the CS coupling generates a   density of topological current  $-\frac{\mu}{m} \mathbf{E}^{\mathrm{m}}_\bot$ for the free field ${\mathbf{A}}$, as well as  a   topological density of charge, $-\frac{\mu}{m} B$,  for the scalar field $\phi_{\mathrm{e}}$.
 
 On the other hand, if we take the magnetic limit, $\phi_{\mathrm{e}}=0$, we have as a consequence $B=0$, what implies $\mathbf{E}^\mathrm{m}_{\perp}=-\partial_t {\mathbf{A}}_{\perp}$ and  $\nabla\cdot {\mathbf{E}}=0$. Using these results we rewrite the equations (\ref{em-5}) and (\ref{em-7}) as:
 \begin{eqnarray}
 	\left[\frac{\nabla^2}{2(m\pm \mu)}-i\partial_t-\frac{\kappa^{2}}{2(m\pm \mu)}\right]A_{\pm}&=&0,   \label{em-10} \\
 	\left(\frac{1}{2m}\nabla^2-i\partial_t-\frac{\kappa^{2}}{2m}\right)\phi_m&=&0.    \label{em-11} 
 \end{eqnarray}
 where $A_\pm=\mathcal A_1\pm i \mathcal A_2$. With these steps, we can see that the CS coupling has made a combination
of the two components of the vector ${\mathbf{A}}$ in order to present them as a free field
$A_\pm$ with a topological mass $m\pm \mu$. Additionally, the component $\phi_{\mathrm{m}}$ behaves like a free scalar field with mass $m$.


\subsection{Second Case: $\tilde{\mu} \neq 0$ and $\mu = 0$}


If we take the  electric limit, $\phi_{\mathrm{m}}=0$, the  equations \eqref{em-5} and \eqref{em-6} are rewritten in the form
\begin{eqnarray}
\left(\frac{1}{2m}\nabla^2-i\partial_t-\frac{\kappa^{2}}{2m}\right)\mathbf{A}&=&-\frac{\tilde{\mu}}{m} \mathbf{E}^{\mathrm{e}}_\bot    \label{em-12} \\
	\left(\frac{1}{2m}\nabla^2-i\partial_t-\frac{\kappa^{2}}{2m}\right)\phi_{\mathrm{e}}&=&0,    \label{em-13}
\end{eqnarray}
From these equations we can see that the CS coupling generates a   topological density of current  $-\frac{\tilde{\mu}}{m} \mathbf{E}^{\mathrm{e}}_\bot $ for the field ${\mathbf{A}}$ and the component $\phi_{\mathrm{e}}$ behaves like a free scalar field with mass $m$.

Finally, taking the magnetic limit, $\phi_e=0$, we obtain for (\ref{em-5}) and (\ref{em-7}) the following equations:
 \begin{eqnarray}
	\left(\frac{\nabla^2}{2m}-\frac{\kappa^{2}}{2m}-\tilde{\mu}\right)A_{\pm}&=&i\partial_t A_\pm,   \label{em-14} \\
	\left(\frac{\nabla^2}{2m}-i\partial_t-\frac{\kappa^{2}}{2m}\right)\phi_m&=&\frac{\tilde{\mu}}{m} B.     \label{em-15} 
\end{eqnarray}
Thus, we can see that the CS coupling has made a combination
of the two components of the vector ${\mathbf{A}}$ in order to present them as a free field
$A_\pm$ with  mass $m$ and an additional constant $\tilde{\mu}$ in its rest energy. Additionally, the CS coupling has also generated a topological density of charge for the free scalar field $\phi_{\mathrm{m}}$.

\section{Podolsky theory with Galilean--symmetry breaking term}\label{Podolsky}

Recently, a relativistic higher derivative version of the Carroll--Field--Jackiw theory has been present in \cite{Ferreira:2024qkd}. The action is given by
\begin{equation}\label{CFJP}
I_{\mathrm{CFJP}} = \int_{\mathcal M} \mathrm{d}^4x\;\left[-\frac{1}{4}F_{\mu\nu}F^{\mu\nu}-\frac{1}{2a^2}\partial_{\mu}F^{\mu\alpha}\partial_{\nu}F_{\;\alpha}^{\nu}-\frac{1}{4}\epsilon^{\alpha\beta\mu\nu}v_\alpha A_\beta F_{\mu\nu}\right].    
\end{equation}
The first two terms in the action are called Podolsky electrodynamics \cite{Podolsky:1942zz} \footnote{It is also called Lee--Wick electrodynamics \cite{Lee:1969fy,Lee:1970iw} by some authors}. The Podolsky term has a second--order derivative term and preserves the gauge symmetry of Maxwell's theory. In the relativistic case, the parameter $a$ is related to the mass of the photon. The last term in \eqref{CFJP} is the usual Lorentz--violating term.

The Galilean covariant version of Podolsky electrodynamics was studied in \cite{Pompeia:2008zz}. Here we propose a Podolsky electrodynamics with Galilean--symmetry breaking term following the same recipe as in Section \ref{Gal_CFJ}. Our action is given by
\begin{equation}\label{podolsky}
I_{\mathrm{Gal-CFJP}}= \int_{\mathcal G}\mathrm{d}^5x\, \left[-\frac{1}{4}\mathcal{F}_{\mu\nu}\mathcal{F}^{\mu\nu}-\frac{1}{2a^{2}}\partial_{\mu}\mathcal{F^{\mu\alpha}\partial}_{\nu}\mathcal{F}_{\;\alpha}^{\nu}-\frac{1}{8}\epsilon^{\alpha\beta\sigma\mu\nu}v_{\alpha\beta}\mathcal{A}_{\sigma}\mathcal{F}_{\mu\nu}\right].    
\end{equation}
At this moment, we do not have an interpretation for the Podolsky parameter $a$ in the non--relativistic theory. The equation of motion is given by
\begin{equation}
\left(1-\frac{\square} {a^2}\right)\partial_{\mu}\mathcal{F^{\mu\sigma}}=\frac{1}{4}\epsilon^{\alpha\beta\mu\nu\sigma}v_{\alpha\beta}\mathcal{F}_{\mu\nu}. 
\end{equation}
where $\square=\partial^\mu\partial_\mu=\nabla^2-2\partial_4\partial_5$. Furthermore, it can be easily shown that the action is invariant under the gauge transformation $\delta \mathcal A_\mu = \partial_\mu \lambda$, where $\lambda$ is the gauge parameter. The canonical analysis of the Podolsky theory \cite{Galvao:1986yq} shows that it is convenient to fix the gauge freedom using the generalized Lorentz gauge:
\begin{equation}
\left(1-\frac{\square}{a^2}\right)\partial_{\mu}\mathcal{A}^{\mu}=0.   
\end{equation}
Then, the equations of motion are reduced to
\begin{equation}
\left(1-\frac{\square}{a^2}\right)\square\mathcal{A}^{\sigma}=\frac{1}{4}\epsilon^{\alpha\beta\mu\nu\sigma}v_{\alpha\beta}\mathcal{F}_{\mu\nu}.  \end{equation}

In order to compare with the results of Section \ref{Gal_CFJ}, let us choose a background of the type \eqref{parameters}. In this case, the electric and magnetic potentials, i.e. the components $\mathcal A_{4,5}$, satisfy the following equation:
\begin{equation}\label{pod_scalar}
\left(a^2-\square\right)\square\phi=0
\end{equation}
where $\phi=(\phi_{\mathrm{e}},\phi_{\mathrm m})$. Since we are interested in massive solutions, we apply the ansatz \eqref{ansatz} and perform the Fourier decomposition. A straightforward computation leads us to the following relation:
\begin{equation}
\left(2m\omega-\mathbf{k}^{2}-a^{2}\right)\left(2m\omega-\mathbf{k}^{2}\right)\hat{\phi}\left(\mathbf{k},\omega\right)=0
\end{equation}
where $\hat\phi$ is the Fourier mode. This equation shows that we have two different dispersion relations which differ from a shift in the rest energy
\begin{equation}\label{two_modes}
\omega = \frac{\mathbf k^2}{2m},\qquad \omega = \frac{\mathbf k^2}{2m}+\frac{a^2}{2m}.
\end{equation}
Therefore, the Podolsky parameter $a$ affects the rest energy and does not couple with $m$, the mass of the photon. This result contrasts with the four--dimensional relativistic theory, where the Podolsky parameter is the mass of the photon. 

We now look at the spatial component of the gauge in the spatial direction. The equation of motion and the generalized Lorentz gauge for the Fourier mode $\hat{\mathbf{A}}(\mathbf{k},\omega)$ are
\begin{eqnarray}
\left(2m\omega-\mathbf{k}^{2}-a^{2}\right)\left(2m\omega-\mathbf{k}^{2}\right)\hat{\mathbf{A}} &=& ina^{2}\mathbf{k}\times\hat{\mathbf{A}},  \\
\left(2m\omega-\mathbf{k}^{2}-a^{2}\right)\mathbf{k}\cdot\hat{\mathbf{A}} &=& 0.
\end{eqnarray}
These equations can be combined to give
\begin{equation}
\left[\left(2m\omega-\mathbf{k}^{2}-a^{2}\right)^{2}\left(2m\omega-\mathbf{k}^{2}\right)^{2}-n^{2}a^{4}\mathbf{k}^{2}\right]\hat{\mathbf{A}}	=	0.
\end{equation}
This is a quartic equation for the frequency. Then, we obtain four frequency modes:
\begin{equation}\label{pod_modes}
\omega	=	\frac{\mathbf{k}^{2}}{2m}+\frac{a^{2}}{4m}\pm\frac{a}{2m}\sqrt{\frac{a^{2}}{4}\pm n\left|\mathbf{k}\right|}.    
\end{equation}
As a test of consistency of our result, let us set $n=0$, i.e., eliminate the dependence on the background. In this case, equation \eqref{pod_modes} reduces to only two modes \eqref{two_modes}. This result should be expected because the equation for the vector field reduces to three copies of \eqref{pod_scalar}, leading to two frequency modes as solutions.

In general, equation \eqref{pod_modes} reinforces that the Podolsky parameter is related to the rest energy. However, the higher derivative term produces splitting in the frequency modes. Also note that the factor $n\left|\mathbf{k}\right|$, which produces birefringence, is present within a square root while the frequency in the Galilean--CFJ \eqref{birefri} is proportional to $n\left|\mathbf{k}\right|$.

Another interesting effect of the dispersion relation \eqref{pod_modes} is the possibility that the term inside the square root is negative, leading to an imaginary frequency, which is a characteristic of dissipative systems. This also hints at the instability of the Podolsky model with Galilean--breaking term.

\section{Final remarks} \label{final_remarks}

Effective electrodynamics has become an important research topic both to investigate physics beyond the standard model and to understand processes that occur in Dirac materials (graphene, topological insulators, Weyl semi--metallic material, etc...). Our focus in this study is to study electromagnetism and higher derivative Podolsky electromagnetism that exhibit anisotropies coming from the LSV contribution described by the Carroll--Field--Jackiw (CFJ) four vector $v_\mu$. The non--relativistic limit is made by the Galilean covariant version of this theory. 

Let us summarize our results. First, we proposed a Galilean covariant version of the Carroll--Field--Jackiw theory \eqref{action}. This model contains the usual Maxwell term, a Proca term, and a Galilean--violating term, with a constant antisymmetric tensor $v_{\mu\nu}$. This tensor can be decomposed as three three--dimensional vectors plus a scalar \eqref{parameters}. For the specific choice where only the scalar parameter survives, we found non--relativistic wave plane solutions that allow two frequency modes, which allow us to describe birefringence. This is one of the main results of our work.

In section \ref{dim_red} we performed the dimensional reduction of the theory, i.e., we considered that the gauge field does not depend on one spatial coordinate, which we choose to be $x^3$. In this scenario, we obtain a planar model, and the Galilean--violating term reduces to two Chern--Simons terms, with couplings $\mu$ and $\bar{\mu}$. In the most general case, where $\mu,\bar{\mu}$ are not zero, the electric and magnetic potentials satisfy a non--relativistic wave equations \eqref{em-6}, \eqref{em-7} only if the (single component) magnetic field $B$ is zero. On the other hand, if we set one of the couplings to zero, the other can be interpreted as a mass term \eqref{em-10} or as a rest energy \eqref{em-14}. 

In section \ref{Podolsky} we proposed a Galilean covariant version of Podolsky electrodynamics with a Galilean symmetry--breaking term. This model also presents the birefringence effect. However, for some specific values of the parameters $a$ (which comes from the Podolsky term) and $n$ (related to the background), the theory becomes unstable, as we can see by checking the frequency modes \eqref{pod_modes}. This is in fact a price we have to pay for dealing with a higher derivative theory. 

As a future perspective, we intend to investigate theories that exhibit VSL without violating CPT symmetry, Euler--Heisenberg electrodynamics, where it appears an effective nonlinear electromagnetic theory stemming from the interaction of photons with virtual electron--positron pairs in a vacuum. It is also worth investigating the equivalence between the Galilean covariant CFJ theory with the model obtained from the large $c$ expansion of the usual Lorentzian CFJ.

\section{Acknowledgements}

H. Belich thanks CNPq for financial support.

\section{Orcid}
\begin{itemize}
\item H. Belich: \url{https://orcid.org/0000-0002-5916-4865}
\item C. Valcarcel: \url{https://orcid.org/0000-0002-5916-4865}
\item E. S. Santos: \url{https://orcid.org/0000-0003-4434-6209}
\end{itemize}

\bibliographystyle{fullsort.bst}

\typeout{}
\bibliography{main} 

\end{document}